\documentclass[11pt,a4paper,dvips]{article}
\usepackage{a4p}
\usepackage{cite,mcite}
\usepackage{subfigure}
\usepackage{atlasphysics}
\usepackage{atlas_title,ifthen}
\usepackage{mathptmx}
\usepackage{amsmath}
\usepackage{helvet}
\usepackage{epsfig}
\usepackage{multirow}
\usepackage{a4wide}
\usepackage{lineno}
\begin{document}
%

\title{Reducing multi-dimensional information into a 1-d histogram}
\author{Mario Campanelli (University College London), William Murray (RAL) }
%
%
\begin{abstract}
We present two methods for reducing multidimensional information to 
one dimension for ease of understand or analysis while maintaining statistical
power. 
While not new, dimensional reduction is not greatly used in high-energy
physics and has applications whenever there is a distinctive feature
(for instance, a mass peak) in one variable
but when signal purity depends on others; so in practice
in most of the areas of physics analysis. 
While both methods presented here assume knowledge of the background,
they differ in the fact that only one of the methods uses a model for the
signal, trading some increase in statistical power for this model dependence. 

\end{abstract}
%
%
%
\section{Introduction}

Statistical analysis in high-energy physics is a complex field, with 
many multi-dimensional analysis techniques used by the various experiments.
The uncomfortable truth remains that while the visualisation
and especially comparison of one dimensional information is relatively
straightforward, two dimensional graphs are already extremely 
difficult to compare  in a meaningful manner and higher dimensional 
distributions are essentially impossible. We therefore present
two techniques intended to simplify that comparison, showing alternative
versions of the projection into the relevant variable, both of
which keep most of
the statistical power of the other hidden ones. 
The basic ideas are not new\cite{ref:splot,ref:aleph}, but they have not been
much used in high-energy physics.

\par

Both techniques allow for fits etc. to be performed in one dimension, with the 
advantages of visibility of the fitting function and the background
distribution, and therefore  easier handling of the
systematics. They thus partially  undo
the `curse of dimensionality'\cite{bellman}.

This paper is divided into different sections.
The first (section~\ref{sec:problem}) 
introduces an example distribution which is  considered throughout this note.
Section~\ref{sec:mario} discusses the
first method, that consists in collapsing a multi-dimensional distribution
into a 1-dimensional one, making only assumptions on the background.
This method can provide a model-independent search for effects like mass
peaks above a Standard Model background assumed to be known. It can also
be used when signal model is uncertain and minimal assumptions 
are to be made.  

The technique shown in section~\ref{sec:bill} also uses a
weighting technique  to reduce a multidimensional distribution into a one
dimensional one, but this time having as an input also the expected signal 
density, such that for un-correlated variables (almost) 
the full statistical power is available in the one-dimensional distribution.
Furthermore the use of  approximate multi-dimensional distributions will
lead to a decrease in the statistical power, and not to false discoveries.

\section{Example search problem }
\label{sec:problem}
\label{sec:estimators}

In order to compare the different methods we introduce an 
example distribution. We chose to consider the simplest case of the reduction 
of a 2-dimensional distribution into 1D, as in this case the full information
can be shown graphically, but the techniques extend trivially into higher
dimensions.
The example chosen is a Gaussian signal in one dimension, decaying exponentially
in the other dimension, while the background is flat in the first dimension
and exponentially falling (sharper than the signal) in the second one:

\begin{eqnarray}
  \sigma_{back} & = & e^{p_T/30}  \\
  \sigma_{sig} & = & e^{p_T/80} \cdot e^{-0.5 ((m-91)/3)^2} 
\label{eq:signal}
\end{eqnarray}

\begin{figure}[htbp]
  \begin{center}
    \includegraphics[width=0.48\linewidth]{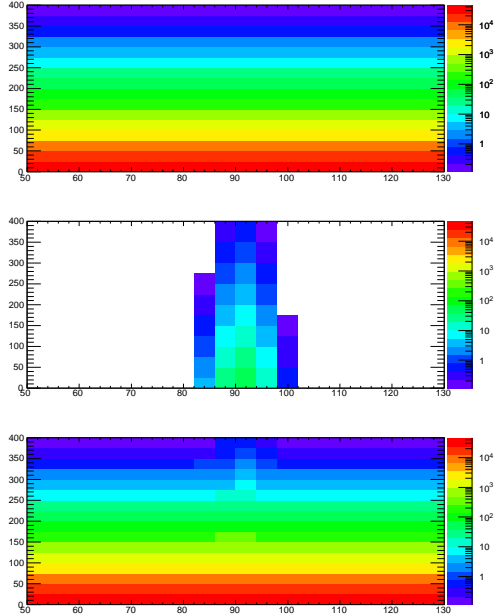}
    \caption{The distribution in mass and \pT\ space of the background (top),
    the signal (centre) and the total (bottom).  }
    \label{fi:distribution}
  \end{center}
\end{figure}

This kind of behaviour may arise, for instance, in the search for a new particle
where the mass and p$_T$ of candidate objects are measured, and an excess is
expected to have a particular mass plus a harder p$_T$ distribution than the
background.
The distribution can be seen in Figure~\ref{fi:distribution}.

%
%
\section{Signal model-independent dimensional reduction}
Let's start by considering the case where we assume we know the distribution
of background, and we want to look for an excess in the data.
We want to reduce the initial N-dimensional distribution into a single
histogram, that keeps the same statistical power as the initial distribution.
We define two N-dimensional histograms, for data and for the expected
background, called $D_{i,j,... p}$ and $B_{i,j,... p}$ and  assume
for simplicity that we want to project them into the first dimension, denoted
by the letter i.
The problem can be seen as finding for each bin $i^\prime$ of the
distribution along
the first variable two values of signal and background that have the same
$\chi^2$ probability as the combination of all the other bins with the 
first variable in bin $i^\prime$.
In practice, for all bins with the first variable in $i^\prime$, we calculate the
low-statistics Poisson approximation\cite{poissonchi2} value of 
\[\chi^2 = 2(B_{i,j,... p}-D_{i,j,... p}) + 2D_{i,j,... p} ln(D_{i,j,... p}/B_{i,j,... p})
\]
and sum these values over all the dimensions to collapse. We also calculate
a signed $\chi^2_{sign}$ distribution, made of a sum the same terms as before, but
with the minus sign for the bins where the number of data events is smaller
than the background expectation.\par 
Then we take the probability P to have the total value of the unsigned 
$\chi^2$ for the
appropriate number of degrees of freedom. Since we want the final 1-dimensional
histogram not to distort the observation from data, we assign the number of
data events for the bin $i^\prime$ to be the total number of events observed there,
i.e. the projection over all other dimensions, with no further manipulation.
For background, if we want to keep the same statistical significance of
the multi-dimensional distribution, we clearly cannot just sum over the other
variables. Instead, we calculate the value of $\chi^2$ that would yield
the probability P for one degree of freedom.
Using the above expression, and given the number of observed data events, there
are two solution for the number of background events, one larger and the
other smaller than the number of data events. We select the first solution
if $\chi^2_{sign}>0$, the second otherwise, in order to properly account for
under-fluctuations in data. 

\label{sec:mario}

%
%
\section{Signal model-dependent dimensional reduction}
\label{sec:bill}

The previous section discussed reduction of a multidimensional space
into a one dimensional one without making any assumption on the distribution
of signal, so it suits general search problems.
The technique here, in contrast, will assume a signal distribution, and is
more suited to the case when a model for the search signal is available.

Similarly to the previous case, the idea is to collapse N variables into a
single one, this time using weights that would optimise the separation between
a known signal and background.

\subsection{Calculation of optimal weighting}

Take the example of a search with two bins, with expected background $b_i$ and
signal $s_i$, and with Gaussian distributed errors on the bins 
$\delta_i $ so  that they each have an expected significance 
$\sigma_i = s_i/\delta_i$.
If we combine them into one bin, applying weights $w_i$  to each of them
as we do so, then the combined total is 

\begin{eqnarray}\label{eq:setup}
b_T &=& W_1 b_1 + W_2 b_2 \\ \nonumber
s_T &=& W_1 s_1 + W_2 s_2 \\ \nonumber
\delta_T^2 &=& (W_1 \delta_1)^2 + (W_2 \delta_2)^2 \\ \nonumber
\sigma_T^2 &=& \frac{(W_1 s_1 + W_2 s_2)^2 }{ (W_1 \sigma_1)^2 + (W_2 \sigma_2)^2} \\ \nonumber
\end{eqnarray}

%
%
%
%
%
%
The total significance is maximised if we use:

\begin{eqnarray}\label{eq:quadratic}
\frac{W_2}{W_1}    & = &   \frac{ \sigma_1^2 } {s_1 } \frac{s_2} {\sigma_2^2  }
\end{eqnarray}

We can set $W_1$ arbitrarily, as the weights are relative. If we 
choose $W_1 = s_1/\sigma_1^2$ then we see  $W_2 = s_2/\sigma_2^2$. 
Given an arbitrary number of bins, and considering merging
them sequentially  pairwise, we see that this weight function
is optimal for any number of bins. There remains however
an arbitrary scale; in this note we scale all weights so that  the
largest  weight in any bin is 1 in all cases.

There is an implicit assumption that the errors are Gaussian.
Note that at this level this
only effects the optimality or otherwise of the combination.
If we assume that errors are of the form 
$\delta_i = \sqrt{b_i}$ then the optimal weight is
$W_i = s_i/b_i$; an alternative  assumption
is $\delta_i = \sqrt{s_i + b_i}$, leading to 
$W_i = s_i/(s_i + b_i)$.
However, if the entire result is to be represented in a single one-dimensional 
plot, which is the aim, then  the errors on the final combined bin must 
 have a Gaussian  form. This in turn requires that the effective number
of events, $(\Sigma_i w_i)^2/\Sigma_i w_i^2$, in each bin must be large.

The question of how to find the signal and background densities to calculate
these weights is not addressed here. However, it must be stressed that
an incorrect weight produces a non-optimal result, and not an incorrect one,
and therefore an approximate method such as the matrix element technique 
or ignoring correlations between variables can legitimately be applied.
The weighting treatment is being applied as a procedure to the data but
all interpretation is reserved for the final distribution.

\section{Application to the example problem}

\begin{figure}[htbp]
  \begin{center}
    \includegraphics[width=0.7\linewidth]{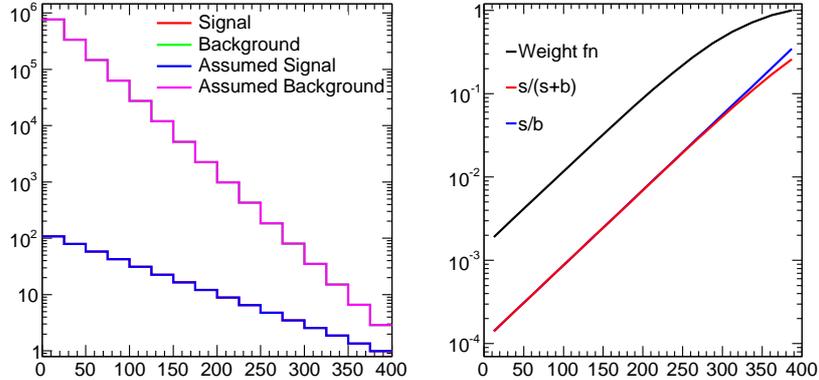}
\vspace*{-0.5cm}
    \caption{The \pT\ spectra. Top is the actual spectra used, bottom
     is the shapes of different possible weight functions. Shown in
     black is the one adopted; the straight blue line shows $s/b$ and 
     $s/(s+b)$ is in red.}
    \label{fi:pt}
  \end{center}
\end{figure}

The problem introduced in section~\ref{sec:problem} is used here.
The \pT\ spectra of signal and background are shown in figure~\ref{fi:pt}.
These functions are perfectly known, and used both to create distributions and 
to fit them.

In the signal weighting reduction method the distinction
between $s/b$ and  $s/(s+b)$ errors needs
discussion.  The derivation assumed errors on the input bins
were all Gaussian, but
in multidimensional distributions this is unlikely to be correct.
The choice of $s/(s+b)$ provides an upper cut-off on the weight 
function which serves to regulate the distribution of weights. This 
allows application of the central limit theorem to the sum of the
weights to calculate the error on the resulting bins.

As has been stressed, it is a  subjective decision
how one builds these weights. For example, the background and signal 
strengths entering in these formula ignore the final dimension (in this 
case, mass). Thus the background strength actually affecting the
analysis on the mass peak is much smaller than is estimated from the other
variables. To reflect this the background strength parameter used to
calculate the weight is  reduced  by width of the  signal divided by the width
of the total distribution.


\begin{figure}[htbp]
  \begin{center}
    \includegraphics[width=0.7\linewidth]{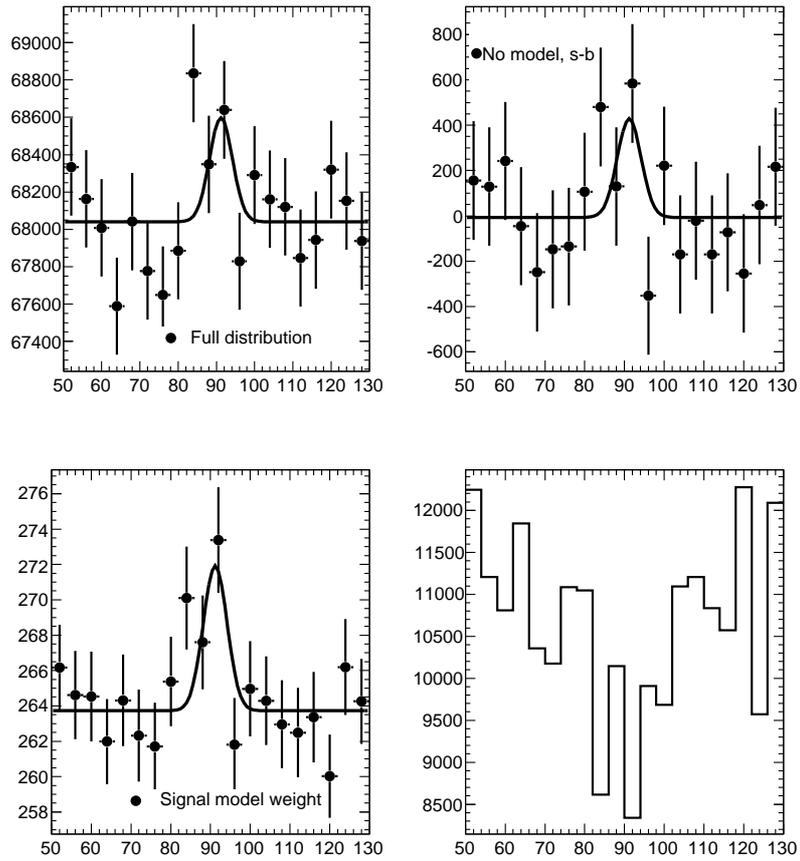}
    \caption{The mass spectra. Top left is all the events projected onto the mass axis. Top right shows the background-subtracted no-model projection.
The bottom plots are the signal model weight projected version on the left
and the corresponding effective number of events per bin on the right.
This is a sample distribution for one trial.
}
    \label{fi:mass}
  \end{center}
\end{figure}

The mass distributions appear as figure~\ref{fi:mass}.
The total projection (top left) has a very poor significance, 
and it is hard by eye to see any signal. The model-independent approach
results in a final
distribution (after background is subtracted from the signal) shown on the 
top right, with a peak which is
more apparent.
The weighted sum, however, (bottom left) also has
a distinctly clearer picture. It will also be seen that it is 
statistically better distinguished as well. The distribution
of the effective number of events (bottom right) is useful to check that 
a sufficiently large number of events is used, to justifying the use of
the central limit theorem.

Four different approaches to the analysis are compared. The first is a 
fit to the full 2D distribution, an the other three are one one dimensional
projections obtained simply by varying a cut or with the reduction methods
discussed. In each case there are two free parameters: the signal level and
the background level.

\begin{figure}[htbp]
  \begin{center}
    \includegraphics[width=0.7\linewidth]{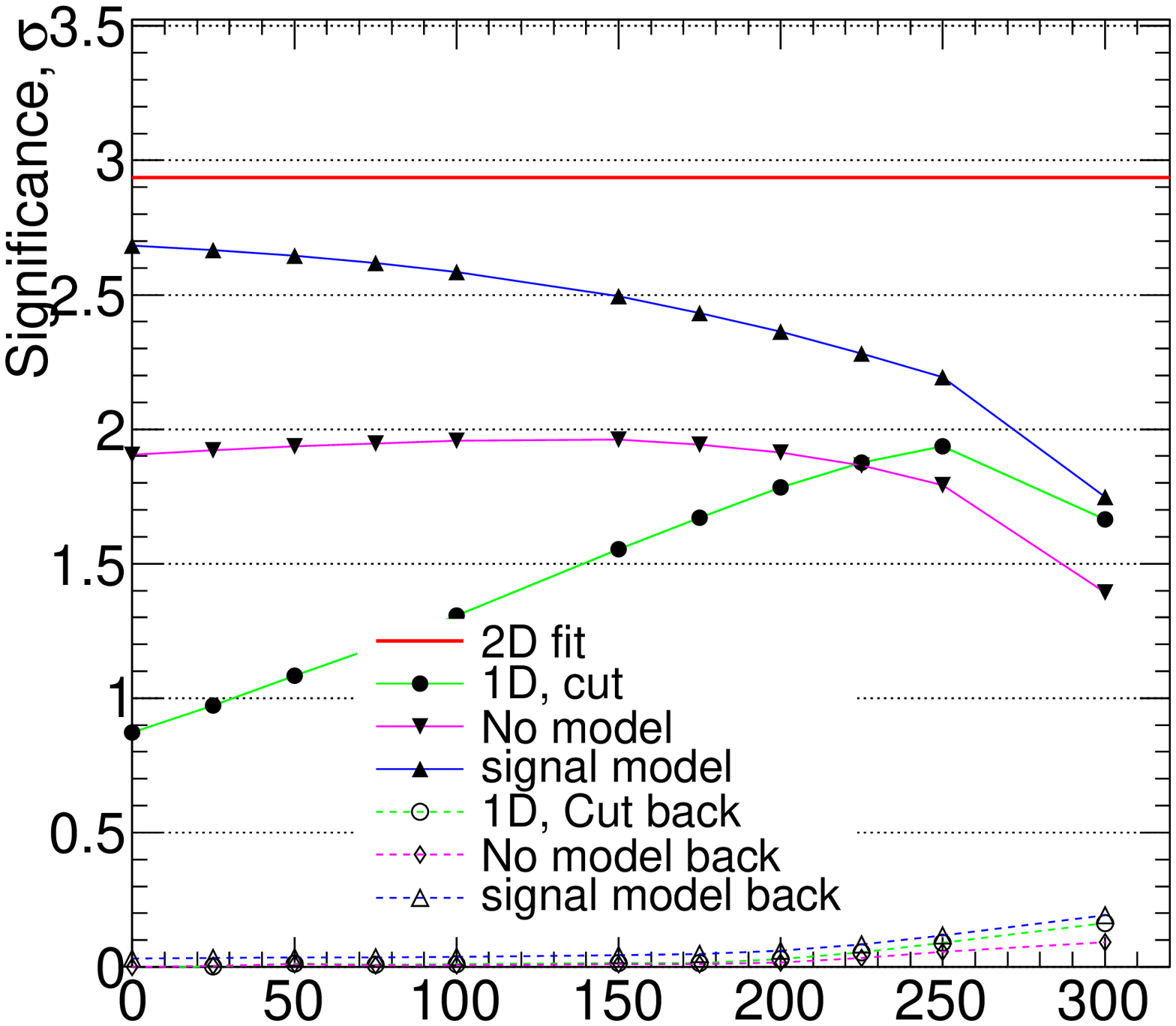}
    \caption{The mean  significance over many trials 
reported by each analysis as a cut on
the \pT\ is varied. In the case of the 2d fit, the cut was not applied
and a fixed line is shown.
Also shown, dashed, are the mean significances reported when fits are made to
background only distributions. This climbs above zero when a $\chi^2$
approximation is used with insufficient events.
}
    \label{fi:v_cut}
  \end{center}
\end{figure}

The statistical power of the various analysis methods is investigated in 
figure~\ref{fi:v_cut}.  For the three analyses where the final
extraction is done with a $\chi^2$ fit in 1D  with the signal and background 
size as free parameters. What is plotted is the claimed significance of the
signal. For the 2D case a likelihood analysis is used as some bins
have low numbers of events. The fit occasionally failed to converge, and so the
signal size has been fixed at the expected value and what is measured
is the difference in log likelihood with and without a signal.
As $-2 \log {\cal L} \simeq \chi^2$, the plot shows the mean of the signed
square-root of twice the  log likelihood, and it will be necessary to test
how accurately this represents the number of sigmas.

The solid curves show the expected significance on
signal plus background samples, while dashed is the expected significance for 
background only. This is not zero because the signal strength
is not allowed to be negative as this causes problem in a likelihood fit.

The most powerful procedure  is to make a 2D fit,
as expected. The signal model weighted fit produces a  performance
almost equal to the 2D fit
when applied to the whole distribution, gradually declining as data is cut away.
The cut based analysis suffers from excessive background when all the data
is considered, and rises in power as this is removed. After some point
the reduction in signal becomes more important and the significance drops.
Note that at this point the expected significance for the background 
only experiments is rising, suggesting that the Gaussian approximation
is breaking down.

\begin{figure}[htbp]
  \begin{center}
    \includegraphics[width=0.48\linewidth]{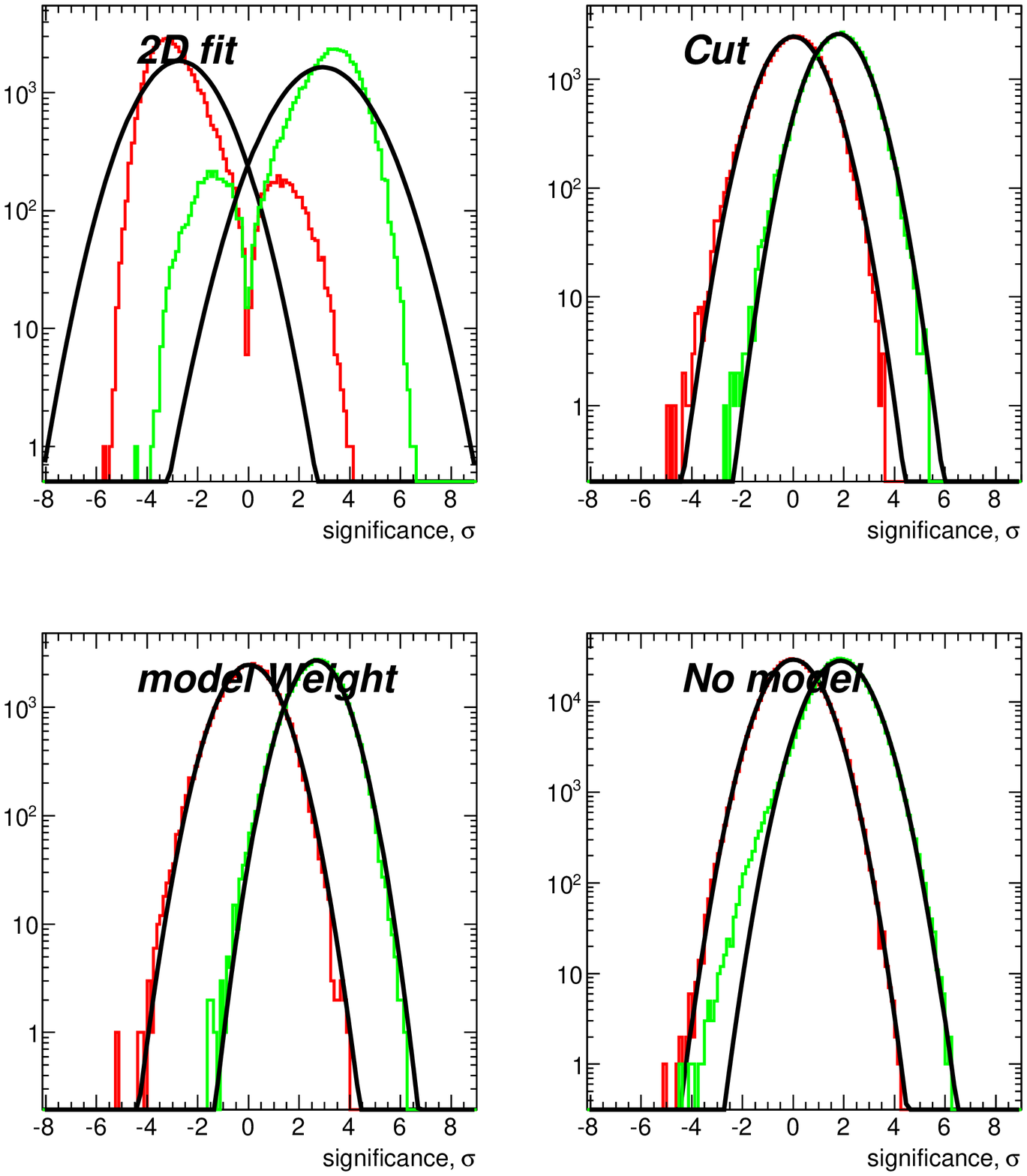}
    \includegraphics[width=0.48\linewidth]{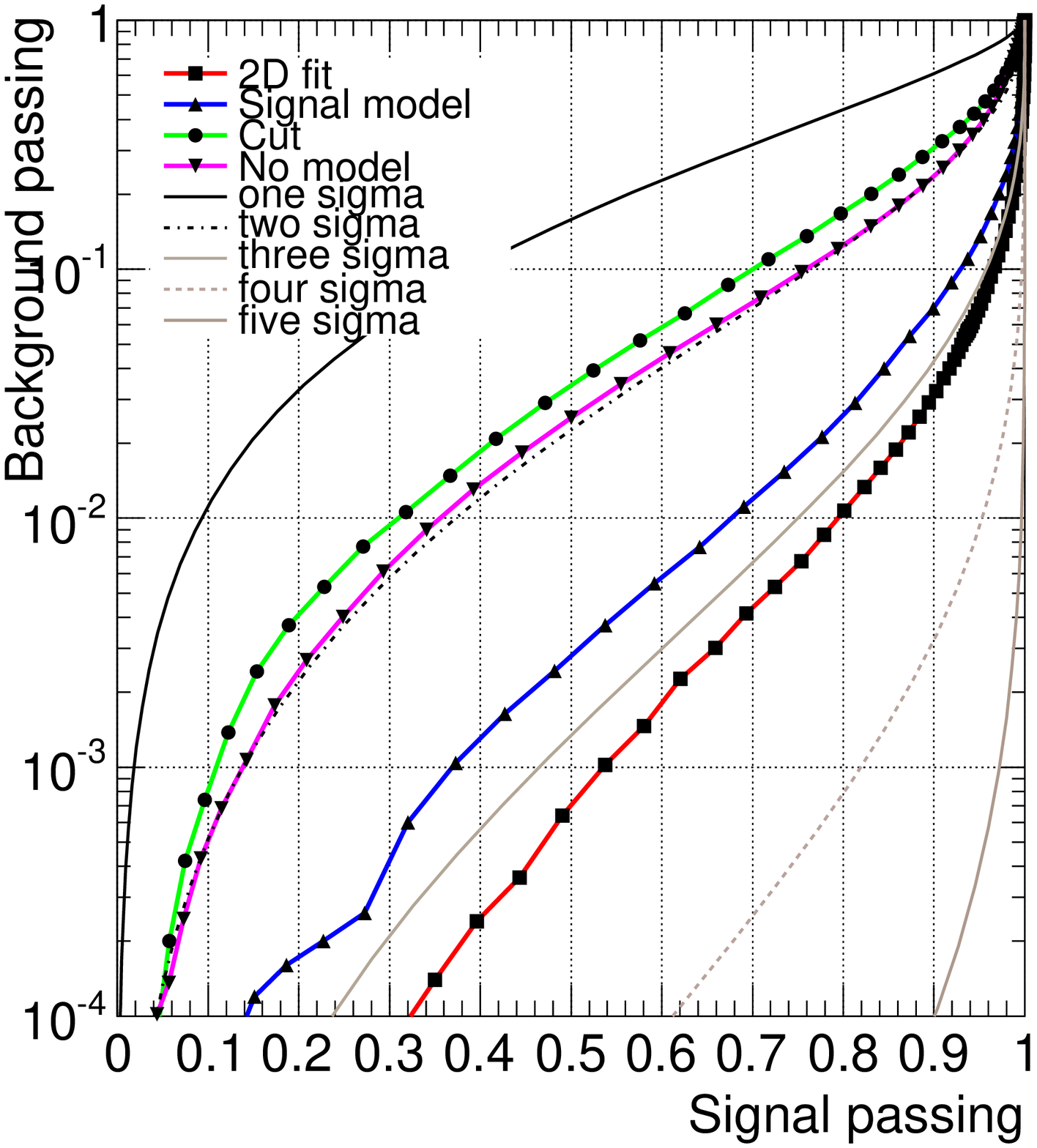}
    \caption{The actual separation power of the fits. The plot on the
left shows the number of sigma reported in toy MC experiments for background
only and signal plus background ensembles.
The plot on the right shows the same data plotted as fractional 
efficiencies which could be achieved by a sliding cut. 
The simple cut based approach is evaluated at a 200~GeV cut, others are for
the full distribution.}
    \label{fi:sep}
  \end{center}
\end{figure}

Figure~\ref{fi:sep} shows the power of each of the fits.
The plots on the left show the significance in toy MC experiments.
The four curves show reasonable agreement with the sensitivities reported
by the fit means, with the only exception being the full 2D fit. The estimated
mean significance was 2.9 $\sigma$, while the observed sensitivity is more
like a 3.1 $\sigma$ separation. This is not surprising - the approximate
connection between  likelihood and $\chi^2$ used is conservative.

\subsection{Signal biases}

The model of the signal used will almost always be, in some details, 
imperfect. How major a problem this  is depends upon the
type of defect, but it will in general 
reduce the sensitivity to a new signal or distort the measurement 
of its parameters. Two biases are consider here: the first is that
the dependence of cross-section on \pT\ has an exponential slope
which is not the one expected, while the second is that there
is a linear correlation between mass and \pT\ in the signal, which 
is assumed not to exist.

\begin{figure}[htbp]
  \begin{center}
    \includegraphics[width=0.48\linewidth]{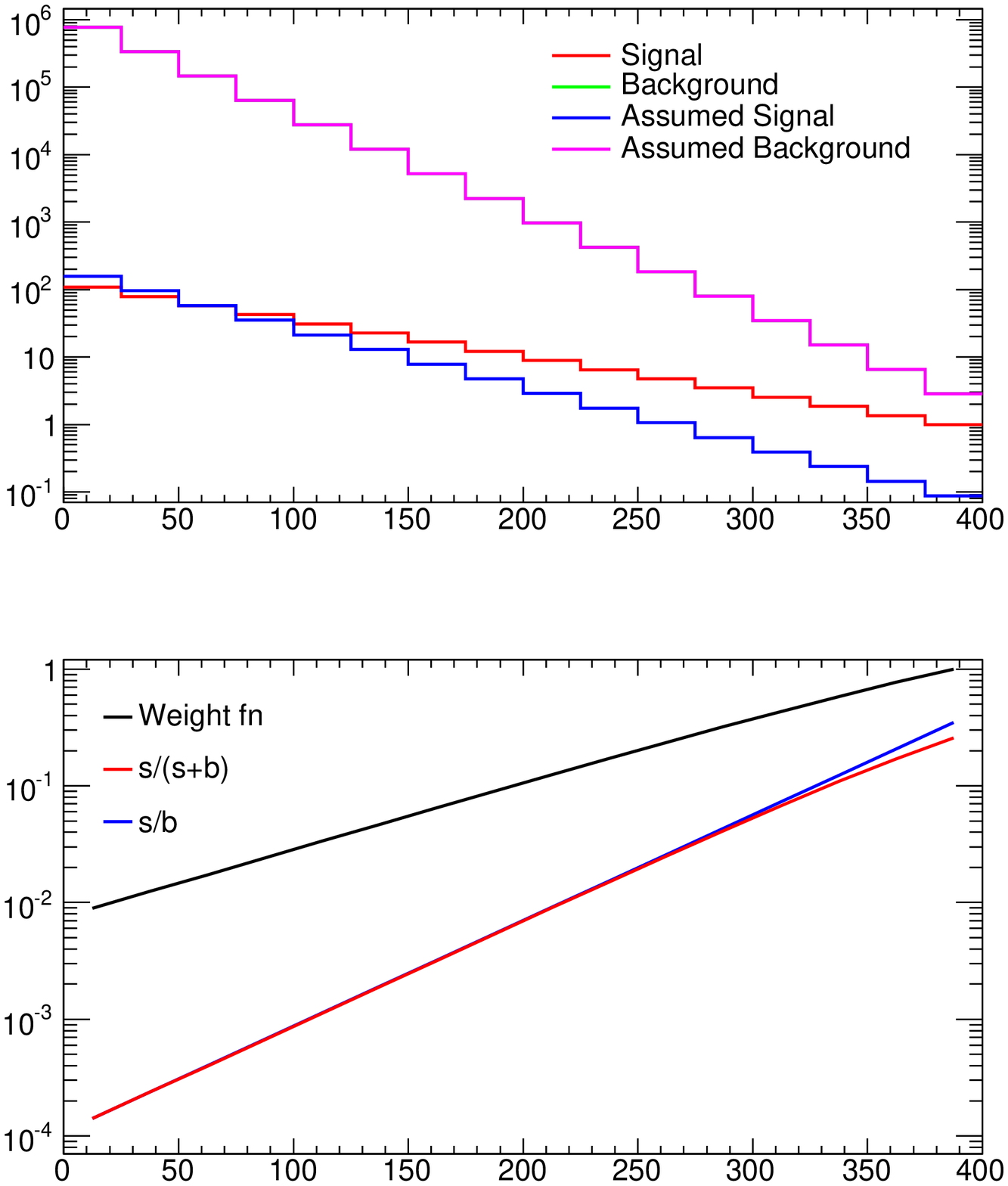}
    \includegraphics[width=0.48\linewidth]{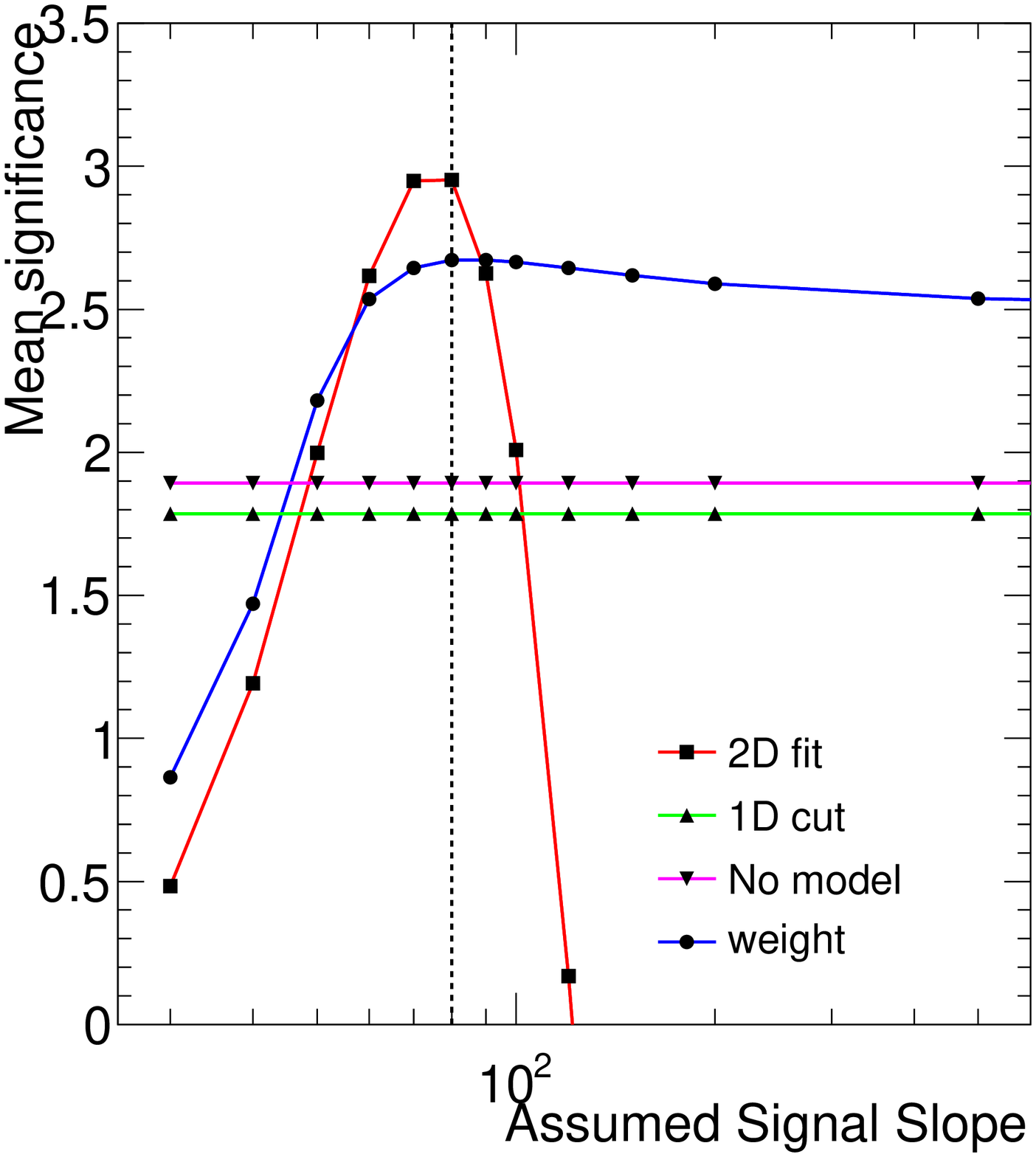}
    \caption{The effect of using the wrong slope in the analysis. The plot on
the left shows an example for an assumed slope of 50. On the right is the
expected significance of the various techniques when the 
wrong signal slope is used in the analysis. All samples have a 
slope of $e^{p_T/80}$, as marked. A large slope parameter corresponds to a 
flat distribution.}
    \label{fi:signalslope}
  \end{center}
\end{figure}

The slope of the signal, given in equation \ref{eq:signal}
as $e^{p_T/80}$ is changed to be 60, 70 100 or 200, while
the 1D and 2D analysis techniques are applied, assuming
a slope of 80. The results for the expected significance are
shown in figure~\ref{fi:signalslope}.

It can be seen that the two 2D fit performs best when the correct
slope parameter is used. However, when too large or too small 
a slope is assumed, the power degrades. The weighted fit shows
similar behaviour, with a lower peak power and a 
 considerably reduced dependence upon the slope value
so that if the slope estimate is 20\% wrong or more it becomes more powerful.
When the assumed slope is 30, this is the same as the background slope;
at this point the weighted analysis is identical to simply ignoring
any $p_T$ dependence.
The cut analysis was not re-optimised but always used a 200~GeV $p_T$ cut,
and therefore does not depend upon the assumed slope; in reality if the
optimisation was done with the wrong slope its performance might reduce.
The no-model approach  has no dependence by construction.

\begin{figure}[htbp]
  \begin{center}
    \includegraphics[width=0.48\linewidth]{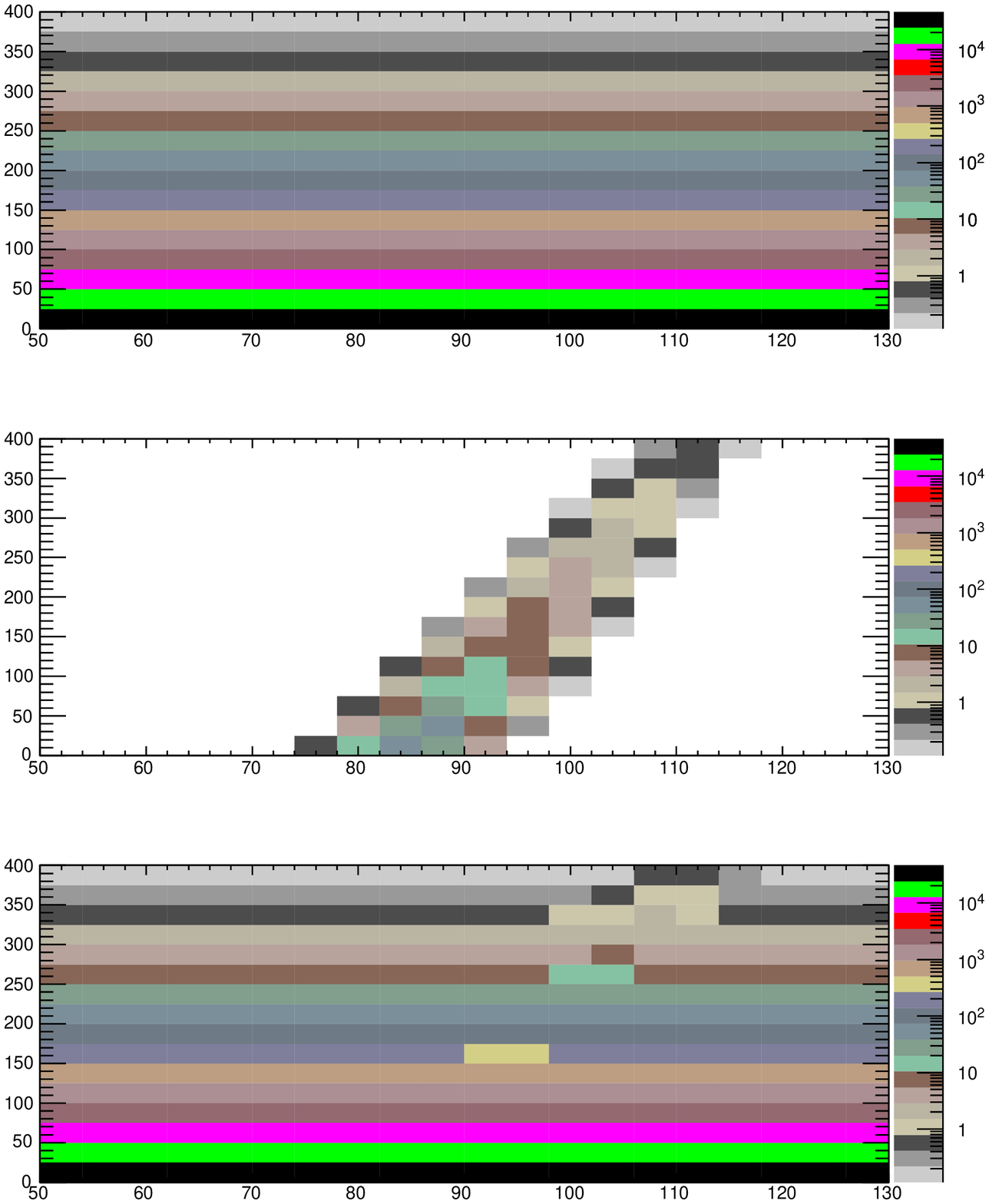}
    \includegraphics[width=0.48\linewidth]{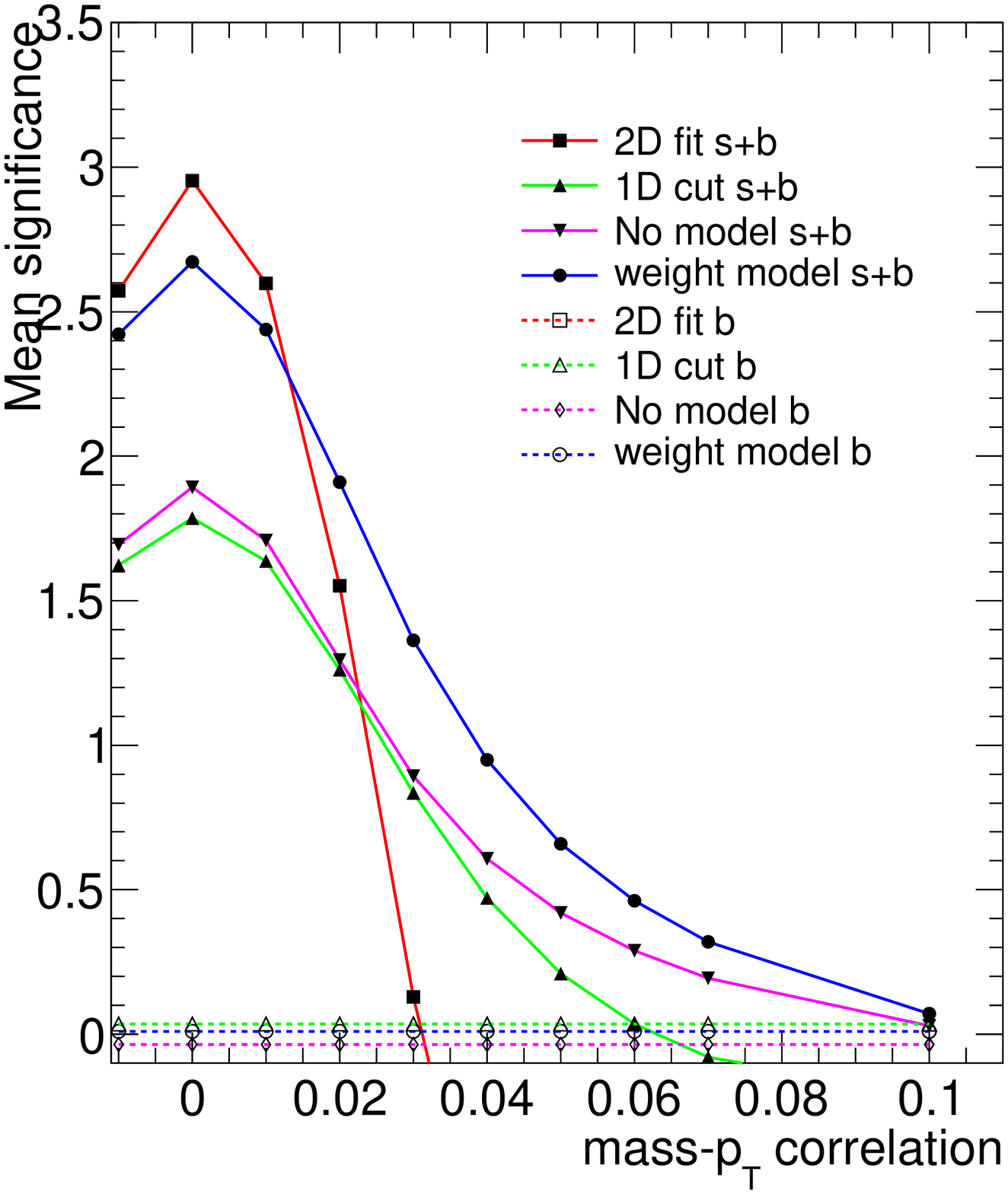}
    \caption{The effect of correlating mass and \pT\ in the signal. Left is the
distribution when a correlation factor 0.7 is used, right is the 
expected significance of various techniques when the 
a correlations between mass and \pT\ is introduced in the signal
but not the analysis. 
}
    \label{fi:signaldmdpt}
  \end{center}
\end{figure}

The second bias investigated is when a correlation is inserted
between the mass and the \pT\ of the signal, with a linear
shift proportional to the \pT.
This means that any any particular \pT\ the signal width in mass 
is unchanged, but that averaged over \pT\ it is widened.
The effects are shown in figure~\ref{fi:signaldmdpt}.

This particular form of distortion means that a correct 2D fit,
allowing for the slope, would not be affected. 
However, all the analyses loose power here, and the 
2D fit is the most sensitive.
This is probably because it is the highest 
weight events, at largest \pT, which move most, thus distorting
the likelihood. The signal weighting method uses weights whose distribution
 is truncated to obtain Gaussian errors, and this seems to protect it in this 
example.

\subsection{Background biases}

\begin{figure}[htbp]
  \begin{center}
    \includegraphics[width=0.48\linewidth]{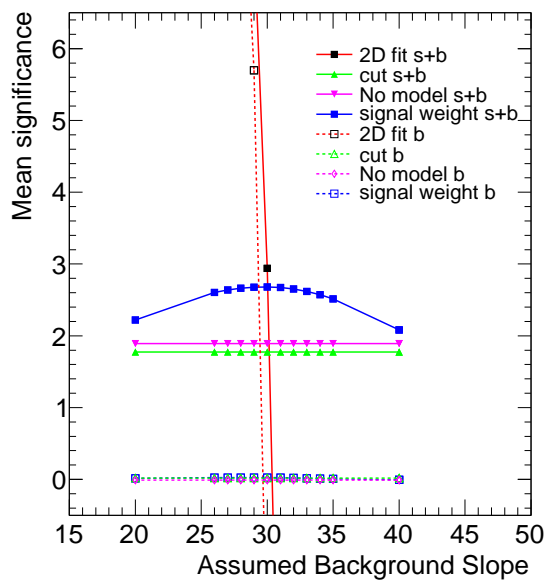}
    \caption{The effect of using the wrong background slope in the analysis.
The
expected significance of the various techniques is shown for background only
and signal plus background samples.
}
    \label{fi:backslope}
  \end{center}
\end{figure}

Figure~\ref{fi:backslope} shows the effect of using the wrong background
slope on the fits. All the fits vary the background level, and for
the one dimensional fits this means that there is no false discovery.
For the 2D
fit the slope error is interpreted as evidence for a signal. 
The blind application of the 2D method clearly produces absurd results.

This is 
a major advantage in using the 1D distributions - the analysis is
done in 1D with less variables and everything visible in simple
histograms. There is, in this case, no doubt enough information
to extract  the slope as part of the 2D fit, but in a real situation
the slope will be more complicated than just an exponential, and
distortions can easily be overlooked.
In the 1D fits  there is essentially no systematic, although the power
of the signal weighting method declines as the wrong background
strength is assumed.

%
%
\section{Conclusions}
Reducing multi-dimensional information to a 1-D histogram can be done in many
different ways. Both methods presented here assume perfect knowledge of the
background, but different in the treatment of signal. One approach has no
assumption whatsoever on the shape of the signal, and aims at looking for
an excess of observed data over the predicted background, weighting more
the events where this excess is more significant (in the form of a smaller
$\chi^2$ probability of being a fluctuation). The other approach assumes 
knowledge of the signal distribution in all variables, and exploits it 
to get the optimal weights for the reduction from N to 1 dimensions.
Results on a simple model show how these reduction methods can be far superior 
to a simple cut-based approach, and have a similar sensitivity to the full
multi-dimensional fit. Since the reduction technique is based on the assumption
of knowing at least the background distribution, we discuss the effects of
systematic biases on our results, concluding that the methods proposed 
here are surprisingly robust against reasonable shape uncertainties

 %
%

%
%


%
%

%
%

\end{document}